\documentclass{llncs}

\usepackage{color}
\usepackage{algorithm}
\usepackage[noend]{algorithmic}
\usepackage{amsmath,tabu}
\usepackage{amssymb,amsfonts,textcomp}
\usepackage{array}
\usepackage{hhline}
\usepackage{caption}
\usepackage{graphicx}
\usepackage{subfig}
\usepackage{verbatim}
\usepackage{booktabs}
\usepackage{url}
\usepackage{pbox}
\usepackage{graphics}
\usepackage{float}
\usepackage{subfloat}
\usepackage{lmodern}

\begin{document}
\title{MaskLink: Efficient Link Discovery for Spatial Relations via Masking Areas}
\titlerunning{LD paper}
\author{Georgios Santipantakis \and Christos Doulkeridis \and George A. Vouros \and Akrivi Vlachou}
\authorrunning{abbreviated author list}
\institute{Department of Digital Systems, University of Piraeus, Greece\\ \email{\{gsant,cdoulk,georgev\}@unipi.gr, avlachou@aueb.gr}}


\maketitle

\begin{abstract}
	In this paper, we study the problem of spatial link discovery (LD), focusing primarily on topological and proximity relations between spatial objects. The problem is timely due to the large number of sources that generate spatial data, including streaming sources (e.g., surveillance of moving objects) but also archival sources (such as static areas of interest). To address the problem of integrating data from such diverse sources, link discovery techniques are required to identify various spatial relations efficiently. Existing approaches typically adopt the filter and refine methodology by exploiting a blocking technique for effective filtering.
	In this paper, we present a new spatial LD technique, called MaskLink, that improves the effectiveness of the filtering step. We show that MaskLink outperforms the state-of-the-art algorithm for link discovery of topological relations, while also addressing some of its limitations, such as applicability for streaming data, low memory requirements, and parallelization. Furthermore, we show that MaskLink can be extended and generalized to the case of proximity-based link discovery, which has not been studied before for spatial data.
	Our empirical study demonstrates the superiority of MaskLink against the state-of-the-art in the case of topological relations, and its performance gain compared to a baseline technique in the case of proximity-based LD.
\end{abstract}

\section{Introduction}
Link discovery (LD) is the process of identifying relations (links) between objects that originate from different data sources $\mathcal{A}$ and $\mathcal{B}$, thereby facilitating several tasks, such as data deduplication and data integration. In the case of spatial data sets, the objective of LD is to discover pairs of spatial objects that satisfy a given relation. Existing works in this area have primarily focused on the discovery of topological relations (within, overlaps, touches, etc.) between spatial objects. 

To avoid the cost of exhaustive comparison ($O(|\mathcal{A}|\cdot|\mathcal{B}|)$) between each pair of objects, blocking techniques~\cite{survey17ld} are typically used that split the 2D space in cells (a process known as grid partitioning or space tiling), assign spatial objects to cells based on overlap, and eventually compare only pairs of objects in each cell. Essentially, the grid cells allow \emph{filtering} of pairs of objects, thus retaining only few candidate pairs of objects, which need to be evaluated in a \emph{refinement} step.

In this paper, we mainly focus on the case of region-to-region relations, and we propose a technique that increases the effectiveness of the filtering step. Our main observation comes from data sources regarding maritime and aviation case studies. 
In such cases, the organization of areas in grid cells typically leaves a significant part of the cell empty. Any area of the other data source that is enclosed in the empty space of a cell cannot produce a topological relation with other areas that overlap with the cell, but the ``disjoint'' relation. Existing techniques would perform comparisons to all areas that overlap with the cell. 

To address this problem, we introduce a technique called MaskLink that explicitly represents the empty space of any grid cell, and exploits it in order to quickly identify objects that would lead to wasteful processing. We apply our technique for link discovery of topological relations between areas, and show that our approach outperforms the state-of-the-art algorithm~\cite{radon17aaai}. In addition, the proposed method can be applied in the case that one source is streaming. More interestingly, we show that MaskLink can be extended to address (a) point-to-region and point-to-point relations, and (b) proximity-based link discovery, thus identifying relations based on distance (e.g., nearby), which are not tackled by existing works.

Concretely, this paper makes the following contributions:
\begin{itemize}
	\item We propose a technique (MaskLink) for LD of topological relations between areas, which improves the efficiency of the filtering step of LD, by eliminating entities that cannot be linked, thereby reducing the number of entities that have to be considered during refinement. 
	\item We show that MaskLink can be extended and generalized, in order to handle relations between other kinds of spatial objects (e.g., points), and proximity-based relations, an issue that has not been studied thus far in the LD literature.
	\item We present the salient features of our approach, which include efficiency, scalability, reduced memory footprint, and parallelization.
	\item We demonstrate the efficiency of MaskLink by comparing against (a) a baseline grid partitioning algorithm, and (b) the state-of-the-art algorithm for region-to-region LD of topological relations, using different real-world data sets.
\end{itemize}

The rest of this paper is structured as follows: Section~\ref{sec:relwork} reviews the related work and Section~\ref{sec:lddefs} presents the necessary notation. Section~\ref{sec:ld} outlines the problem of LD for topological relations and presents MaskLink. Then, in Section~\ref{sec:ldprox}, we show how MaskLink can be extended and generalized towards other types of spatial objects and proximity-based relations. Section~\ref{evaluation} presents the findings of our comparative, empirical study, and Section~\ref{sec:concl} concludes the paper.

\section{Related Work} \label{sec:relwork}

Link discovery is a challenging topic which relates to record linkage~\cite{DBLP:journals/tkde/Christen12}, deduplication~\cite{DBLP:journals/tkde/ElmagarmidIV07}, and data fusion~\cite{DBLP:journals/csur/BleiholderN08}.
However, even though the topic of link discovery has attracted much interest and attention lately (see~\cite{survey17ld} for a recent survey), there is not much work on the challenging topic of topological and proximity link discovery. This paper focuses on these topics, 
towards the discovery of relations. In the next paragraphs, we provide an overview of related link discovery frameworks and techniques. 

LIMES is a generic link discovery framework~\cite{limes11ijcai} for metric spaces that uses the triangular inequality in order to avoid processing all possible pairs of objects. For this purpose, it employs the concept of exemplars, which are used to represent areas in the multidimensional space, and tries to prune entire areas (and the respective enclosed entities) from consideration during the refinement step. Another link discovery framework is SILK~\cite{silk11webdb}, proposing a novel blocking method called MultiBlock, which uses a multidimensional index in which similar objects are located near to each other. In each dimension the entities are indexed by a different property or different similarity measure. Then, the indices are combined together to form a multidimensional index, which is able to prune more entities by taking into account the combination of dimensions. 

HR$^3$~\cite{hr312iswc} and HYPPO~\cite{DBLP:conf/semweb/Ngomo11} address link discovery tasks when the property values that are to be compared are expressed in an affine space with a Minkowski distance. Both approaches are designed with main objectives to be efficient and lossless. In addition, HR$^3$~\cite{hr312iswc} comes with theoretical guarantees on reduction ratio, a metric that corresponds to the percentage of the Cartesian product of two data sources that was not explored before reporting the link discovery results. However, all aforementioned approaches for LD do not explicitly focus on spatio-temporal link discovery, nor do they tackle the streaming nature of data sources. 

The spatial link discovery methods~\cite{DBLP:conf/semweb/Ngomo13,radon17aaai} apply grid partitioning on the input data sources, in order to perform efficiently the \emph{filtering step}, i.e., avoid comparisons between entities of the input data sources, that will not result to a link association. Then, the \emph{refinement step} follows, where different optimizations are employed in order to minimize the number of computations necessary to produce the correct result set. RADON~\cite{radon17aaai} is the most recent approach for discovering topological relations between data sources of areas, and can discover efficiently multiple relations using space tiling. One of its main techniques for efficiency relies on the use of caching to avoid recomputing distances. However this imposes non-negligible requirements for main memory, especially for large data sources. Furthermore, RADON employs an optimization based on minimum bounding boxes (MBBs) for deciding the granularity of the grid. This means that a data source providing point geometries (e.g. for touches or within relations) would force the construction of an ``infinite'' number of cells. This indicates that RADON cannot handle point-to-region topological relations. We present RADON in more detail in Section~\ref{sub:radon}.

ORCHID~\cite{DBLP:conf/semweb/Ngomo13} is another grid partitioning method, which studies the problem of discovering all pairs of polygons, such that their Hausdorff distance (practically Max-Min distance) is below a given threshold. It also employs space tiling to improve the filtering step. Also, it employs bounding circles as approximations of polygons together with applying the triangular inequality and already computed distances to avoid computing new distances, thus pruning areas without distance computations. Smeros et al.~\cite{smeros16ldow} study link discovery on spatio-temporal RDF data. The authors study several topological relations that are defined on polygons. The topological relations do not take into account proximity nor distance of the polygon, and several of those relations are meaningful only when both data sources include polygons and not points. The algorithm provided creates an equi-grid, and filters out cells that contain polygons that cannot satisfy the relation. 


In summary, none of the above papers targets both topological and proximity relations, which is a distinguishing feature of this work.
Also, it is noteworthy that we have successfully applied our technique in the case of real-time link discovery of spatial relations in the maritime domain, in order to
enhance complex activity recognition~\cite{swj2017subm}. 


\section{Notation and Definitions} \label{sec:lddefs}

In this section, we provide the necessary definitions and problem setting for topological and proximity-based link discovery. 

Let $\mathcal{A}, \mathcal{B}$ two data sources of spatial representations of entities. Each $A \in \mathcal{A}$ (resp. $B \in \mathcal{B}$) is represented as a set of points $a_i$, i.e., $A=\{a_1,a_2,\dots,a_n\},~n\geq 1$, and we write $a_i \in A$ to denote that $a_i$ is included in the representation of $A$. Let $d(p,p')$ denote the distance between two points $(p.x,p.y)$ and $(p'.x,p'.y)$ in 2D. Without loss of generality, we employ the Euclidean distance function $d(p,p')=\sqrt{(p.x-p'.x)^2+(p.y-p'.y)^2}$ to quantify the spatial distance of two points. Other distance functions, such as the Haversine distance (a.k.a. orthodromic distance) \cite{heavMath}, are also applicable. Also, note that for small distances the Euclidean distance serves as an approximation of the Haversine distance, and the relative error is small. Further, overloading the distance function, we denote $d(A,B)$ the distance between two entities $A,B$ such that $d(A,B)=\min_{a \in A, b \in B} d(a,b)$, where $a$ (resp. $b$) and be can be any point in $A$ (resp. $B$) or any point in the line segments between consecutive points $(a_i,a_{i+1})$,with $a_i, a_{i+1} \in A$ (resp., consecutive points $(b_i,b_{i+1})$,with $b_i, b_{i+1} \in B$).


Given the above notations, the following relations can be discovered by the link discovery process:

\begin{definition} $\mathit{within(A,B)}$ or $\mathit{covers(A,B)}$: Given the spatial representations $A \in \mathcal{A}, B \in \mathcal{B}$, 
	$\mathit{within(A,B)}$ is true, if $A$ is enclosed in $B$, i.e. the intersection of $A,B$ is $A$. 
\end{definition}

\begin{definition} $\mathit{nearby(A,B,\theta)}$: Given the spatial representations $A \in \mathcal{A}, B \in \mathcal{B}$, and a distance threshold $\theta$, $\mathit{nearby(A,B,\theta)}$ is true, if $d(A,B) \leq \theta$.
\end{definition}

\begin{definition} $\mathit{disjoint(A,B)}$: Given the spatial representations $A \in \mathcal{A}, B \in \mathcal{B}$, 
	$\mathit{disjoint(A,B)}$ is true, if the intersection between $A$ and $B$ is empty. 
\end{definition}

\begin{definition} $\mathit{meet(A,B)}$: Given the spatial representations $A \in \mathcal{A}, B \in \mathcal{B}$, 
	$\mathit{meet(A,B)}$ is true, if the intersection of boundaries of $A,B$ is not empty, and the intersection between the interiors of $A,B$ is empty. 
\end{definition}

\begin{definition} $\mathit{overlap(A,B)}$: Given the spatial representations $A \in \mathcal{A}, B \in \mathcal{B}$, 
	$\mathit{overlap(A,B)}$ is true for $A,B$, if the intersection of their boundaries and the intersection of their interiors are both not empty. 
\end{definition}

Subsequently, we denote by $area(X)$  all the points enclosed in an area of a spatial entity $X$, including the boundary. Also, $Overlap\_S(A,B)$ is the area in the intersection of A and B. Formally, given an area $B$, in case $A$ is an area and holds that $overlap(A,B)=True$, then $Overlap\_S(A,B)=area(A) \cap area(B)$. In case $A$ is a point, then $Overlap\_S(A,B)=A$.

It is trivial to show that if two geometries $A,B$ are not disjoint, then at least one of the topological relations holds. 



\section{Link Discovery of Topological Relations between Regions} \label{sec:ld}

In this section, we present solutions for link discovery of topological relations between areas, focusing first on a baseline approach (Section~\ref{sub:bsl}), and then on the state-of-the-art algorithm RADON~\cite{radon17aaai} (Section~\ref{sub:radon}). To the best of our knowledge, RADON currently outperforms other link discovery approaches for topological relations. Then, in section \ref{masklink} we present our proposal for efficiently filtering candidate pairs of entities, MaskLink. 

\subsection{A Baseline Technique}\label{sub:bsl}

Given the \emph{target} $\mathcal{B}$, and \emph{source} $\mathcal{A}$ data sources, and a relation $r$ in the set of topological relations $R$ defined above, the goal of link discovery is to detect the pairs $(\sigma,\tau) \subseteq \mathcal{A} \times \mathcal{B}$, where $\sigma \in \mathcal{A}$ and $\tau \in \mathcal{B}$, s.t. $(\sigma,\tau)$ satisfies $r$. 

A brute force link discovery algorithm would have to perform the geometrical test between $A$ and any entity in $\mathcal{B}$, thereby producing the result using 
$O(n*m)$ comparisons, where 
$n=|\mathcal{B}|,m=|R|$. 
However, this cost may be prohibitively expensive  in practice for large data sets.

To avoid this excessive cost, blocking techniques are typically employed in the filtering step, in order to prune the candidate pairs of entities considered during the refinement step. In the case of link discovery for spatial data, the prevalent blocking mechanism is to apply \emph{grid partitioning} of the 2D space (also known as \emph{space tiling}~\cite{DBLP:conf/semweb/Ngomo13}). Essentially, the space is partitioned to cells, and any entity is assigned to 
the cells that include this or any of its parts (e.g. in the case of entities represented by points, we need to consider the cell that includes the point, and in case of areas, we need to consider the set of overlapping cells).

We say that a spatial representation $A$ is \textit{assigned} to a cell $c$, if $c$ includes or overlaps with $A$. Then, to compute each relation $r \in R$, the spatial representation $A$ is compared only against those entities in $\mathcal{B}$ assigned to the cells where $A$ is assigned. This approach results to $O(k*m)$ comparisons with $k=D(A)$ and $m=|R|$, where $D(A)=\{B| B \in \mathcal{B}$ and assigned to any cell where $A$ is also assigned$\}$ 
(typically $k\ll n$). We refer to this link discovery technique as \emph{baseline}. 

\subsection{RADON: A State-of-the-Art Technique}\label{sub:radon}
RADON~\cite{radon17aaai} is the state-of-the-art algorithm for topological link discovery between areas  archived in fixed data sources. Its basic operation relies on the use of \emph{space tiling}. In the 2D spatial domain, this is essentially an equi-grid partitioning of the space, where an arbitrary number of rectangular cells are created. As a result, RADON relies on the same baseline (admittedly with optimizations): first, it organizes the target data source $\mathcal{B}$ in the grid, by assigning minimum bounding boxes (MBBs) of areas to cells, and then it processes each area of $\mathcal{A}$ by assigning also to overlapping cells. 

To achieve improved performance, RADON utilizes the following techniques. First, it selects the data set that will be organized in the cells, based on a heuristic, thus offering performance gain at runtime (this is called \emph{swapping}). Second, it only assigns areas $A \in \mathcal{A}$ to a cell that already contains areas from data set $\mathcal{B}$. This is mentioned as sparse space tiling. Third, it applies a caching mechanism in order to avoid re-computing relations $(A,B)$ that have been previously computed, e.g., due to $A, B$ spanning multiple cells.

Despite the benefits in efficiency offered by these techniques, their adoption also limits the applicability of RADON. For instance, swapping requires that both data sets are available beforehand, thereby rendering this technique (and consequently RADON) inapplicable in the case of streaming data sources. Also, RADON requires both data sets to be stored in-memory, and in combination with the caching mechanism, it adds a significant overhead in main memory. In contrast, in our approach, only the target data source $\mathcal{B}$ is organized in memory, thus offering scalability regardless of the size of $\mathcal{A}$, which can be a streaming source, as well. Motivated by these shortcomings, we present our technique, called MaskLink, in the following subsection.

\subsection{Link Discovery with MaskLink}\label{masklink}


In this section we present the MaskLink technique in the case where at least the target data source includes entities whose spatial representation are areas. 

In several cases of spatial link discovery, grid cells contain a significant amount of ``empty space'', namely the part of the cell that overlaps with no areas of the target data set $\mathcal{B}$. 
Consider a spatial representation $A \in \mathcal{A}$ of an entity that overlaps with a cell, and $k$ spatial representations $\{B_1,\dots,B_k\} \in \mathcal{B}$ that also overlap with the given cell.
Our observation is that if $A$ is disjoint to all the spatial representations in $\{B_1,\dots,B_k\}$ in this cell, then we can safely infer that there are no other (except ``disjointness'') topological relations to be discovered in this cell between $A$ and any area in $\{B_1,\dots,B_k\}$. 

Motivated by this observation, we propose the masking technique to explicitly represent the empty space within cells as yet another area. Thus, for each grid cell $c$, we construct an artificial area called \emph{mask} of $c$, which is defined as the difference between the cell and the union of areas overlapping with the cell, i.e.,

$mask(c) = area(c) - (area(c) \cap \bigcup_i area(B_i))$.

\begin{figure}[thb]
	\centering
	\includegraphics[width=0.4\textwidth]{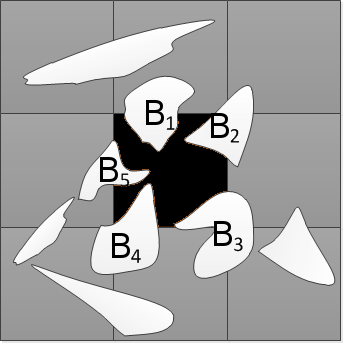}
	\caption{Illustration of the mask of a grid cell that overlaps with five areas $B_1,\dots,B_5$ 
		.}
	\label{fig:mask}
\end{figure}

Figure~\ref{fig:mask} shows an example of the mask of a cell; the middle cell overlaps with areas in $\{B_1,\dots,B_5\}$, and the mask of the cell is the area represented in black color.

Having the mask of a cell as yet another area, we can devise an efficient algorithm for link discovery that eagerly avoids comparisons to geometries for spatial representations enclosed in the mask of a cell. In practice, after we identify a cell $c$ to which the spatial representation $A$ is assigned, and in case $A$ is an area, we first construct the $Overlap\_S(c,A)$. We then compare $Overlap\_S(c,A)$  to the mask of the cell, to check if it is enclosed in the empty space. If this single comparison returns true, we stop processing this entity, thereby saving $k$ comparisons. For the typical case where a cell contains several areas, this technique  prunes several candidate pairs of entities, saving computational time in the refinement step of the LD process.

\begin{algorithm}[t]
	\caption{Spatial LD algorithm for topological relations using mask.}
	\label{alg:topol}
	\begin{algorithmic}[1]
		\STATE \textbf{Input:} Grid cells $C=\{c_1,\dots,c_m\}$, spatial representations $\mathcal{B}=\{B_1,\dots,B_n\}$, $A$
		\STATE \textbf{Output:} Set of relations $r(A,B_i)$ s.t. $r$ is a topological relations and $B_i \in \mathcal{B}$
		\STATE \textbf{Requires:} Grid has been constructed, and spatial representations in $\mathcal{B}$ have been assigned to overlapping cells
		\STATE $\mathcal{B}^w \leftarrow \emptyset$
		\STATE locate cells $\Psi \subseteq C$ that enclose or overlap with $A$\label{lin:1} 
		\FOR{each $c_i \in \Psi$}
		\IF{within($Overlap\_S(c_i,A)$, $mask(c_i))$}	\label{lin:2}
		\FOR{each $B_j \in c_i$}
		\STATE $\mathcal{B}^w \leftarrow \mathcal{B}^w \cup \{disjoint(A,B_j)\}$
		\ENDFOR
		\STATE \textbf{return} $\mathcal{B}^w$
		\ELSE 
		\FOR{each $B_j \text{ assigned to } c_i$} \label{lin:3}
		\FOR{each relation $r$}		
		\IF{r($A$, $B_j$)} 
		\STATE $\mathcal{B}^w \leftarrow \mathcal{B}^w \cup \{r(A,B_j)\}$  \label{lin:4}
		\ENDIF
		\ENDFOR
		\ENDFOR
		\ENDIF
		\ENDFOR
		\STATE \textbf{return} $\mathcal{B}^w$
	\end{algorithmic}
\end{algorithm}

Algorithm~\ref{alg:topol} presents the pseudo-code for discovering any topological relation link between the spatial representation $A$ and the spatial representations of $\mathcal{B}$. As a prerequisite, the grid has already been constructed and the spatial representations of $\mathcal{B}$ have been assigned to cells. This is essentially a pre-processing step. In the first step, the cells $\Psi$ to which $A$ is assigned are determined (line~\ref{lin:1}). This operation is performed in constant time $O(1)$ in the case of equi-grids. Then, for each $c_i \in \Psi$, the algorithm checks if $Overlap\_S(c_i,A)$ is enclosed in the mask $mask(c_i)$ of cell $c_i$ (line~\ref{lin:2}). In the latter case, no further processing is required, and the algorithm terminates returning the inferred set of disjoint relations to the spatial representations assigned to $c_i$. If it is not contained, then each relation against all areas $B_j$ in cell $c_i$ (line~\ref{lin:3}) is checked. For those areas $B_j$ that satisfy a relation $r(A,B_j$), we append  the discovered links in the result set $\mathcal{B}^w$ (line~\ref{lin:4}), and return $\mathcal{B}^w$. 

The lines~\ref{lin:3}--\ref{lin:4} of Algorithm~\ref{alg:topol} are processed in parallel, i.e., each iteration in the for loop is 
carried out by a different 
thread (`worker'). The number of concurrent workers is usually a predefined constant w.r.t. system configuration, to allow uninterruptible system operation (in our experiments we employ 4 workers). We have enabled multi-thread processing using a \emph{pool of tasks}, populated with the refinement tasks of $r(A,B_j)$. As soon as a worker is available and the pool contains tasks, the next task is selected and assigned to the worker for processing.

Also, the algorithm is amenable to parallelization beyond the scope of a single machine (e.g. for high velocity streams), by simply partitioning the target data source of spatial representations to the available processing nodes, each of which runs an instance of Algorithm~\ref{alg:topol}. 


\subsubsection{Cost Analysis}

To quantify the gain offered by MaskLink as well as the conditions under which this gain is achieved, we perform a simplified back-of-the-envelope computation.

For a given cell $c$, let $k$ denote the number of areas $\{B_1,\dots,B_k\} \in \mathcal{B}$ included in $c$. Further, given an area $A \in \mathcal{A}$ that overlaps with $c$, let $p$ denote the probability that $A$ is enclosed in the mask of $c$, i.e., $within(A,mask(c))$ is true. The average number of comparisons performed by MaskLink for an area $A \in \mathcal{A}$ is then: $p + (1-p)(k+1)$, since with probability $p$ only one comparison is performed (between $A$ and $mask(c)$), whereas with probability $1-p$ we perform $k+1$ comparisons (where $k$ refers to the comparisons between $A$ and the $k$ $B_i$). In contrast, an algorithm such as baseline would perform $k$ comparisons. Thus, the gain achieved  by MaskLink is quantified as: $\frac{p + (1-p)(k+1)}{k}$ and must be smaller than 1. If we solve this inequality, it turns out that (for a given a cell) the condition in order to have gain is that: $p > 1/k$.

\section{Extending and Generalizing MaskLink} \label{sec:ldprox}

In this section, we first show that the MaskLink technique can be extended to support other types of relations besides topological relations, such as proximity relations. Then, we generalize the proposed technique to handle spatial link discovery tasks for other types of spatial objects, instead of only areas.

\subsection{Extending MaskLink for Proximity Relations}

Interestingly, the MaskLink technique is applicable also for link discovery of proximity relations between spatial representations of entities in $\mathcal{A}$ and $\mathcal{B}$. More concretely, let us consider the `nearby' relation defined in Section~\ref{sec:lddefs}.
Recall that the `nearby' relation is defined using a spatial threshold $\theta$ and returns true when it holds that $dist(A,B) \leq \theta$, for two areas $A, B$. Proximity link discovery concerns the identification of all such relations between entities in data sets $\mathcal{A}$
and $\mathcal{B}$. In more detail, given the spatial representation of any entity (either a point or an area) $A \in \mathcal{A}$, we wish to discover the subset of spatial representations  in $\mathcal{B}$ that are located at most at distance $\theta$ from $A$. 

Let $B'=buff(B,\theta)$ denote the expanded area of $B$ that contains all points in space located  in distance less or equal to $\theta$ from any  point in $B$, and can be computed using a standard library for computational geometry. Any such area, depending on threshold $\theta$, is called $\theta$-buffered  area.

Assuming that $A$ is an area, we make the following basic observation: if $A,buff(B,\theta)$ are disjoint, then $nearby(A,B,\theta)=false$. We extend this observation for more than one area $B$, to make it applicable for cells. 

More formally, given a cell $c$ that overlaps with the area $A \in \mathcal{A}$ and $k$ areas $\{B_1,\dots,B_k\} \subseteq \mathcal{B}$, if $Overlap\_S(c,A)$ is disjoint to any $buff(B_i,\theta)$, 
then we can infer that $nearby(A,B_i,\theta)=false$ for all $B_i$, $1 \leq i \leq k$.  A similar case holds in case A is a point: if $A$ is not enclosed to any $buff(B_i,\theta)$, then we can infer that $nearby(A,B_i,\theta)=false$ for all $B_i$, $1 \leq i \leq k$.

Based on these observations, we slightly adjust MaskLink for the relation `nearby'. The main adjustment concerns the way the mask of each cell is computed. First, we expand each area $B_i$ by $\theta$, and then the mask  of the cell is computed as previously, using the $\theta$-buffered areas instead of the actual areas $B_i$. To differentiate this mask of a cell $c$ from the one used in the previous algorithm, we denote it by $mask^{\theta}(c)$.

\begin{algorithm}[t]
	\caption{Spatial link discovery algorithm for relation \textit{`nearby'} using mask.}
	\label{alg:nearby}
	\begin{algorithmic}[1]
		\STATE \textbf{Input:} Grid cells $C=\{c_1,\dots,c_m\}$, Areas $\mathcal{B}=\{B_1,\dots,B_n\}$, $A$, threshold $\theta$ 
		\STATE \textbf{Output:} Set of relations $r(A,B_j)$ s.t. $r\in\{nearby,disjoint\}$ and $B_j \in \mathcal{B}$
		\STATE \textbf{Requires:} Grid has been constructed and areas have been assigned to  cells 
		\STATE $\mathcal{B}^n \leftarrow \emptyset$
		\STATE locate cells $\Psi \subseteq C$  to which $A$ is assigned \label{lin:1n} 
		\FOR{each $c_i \in \Psi$}
		\IF{ $within(Overlap\_S(c_i,A),mask^{\theta}(c_i))$}	\label{lin:2n}
		\FOR{each $B_j \in c_i$}
		\STATE $\mathcal{B}^n \leftarrow \mathcal{B}^n \cup \{disjoint(A,B_j)\}$
		\ENDFOR
		\STATE \textbf{return} $\mathcal{B}^n$
		\ELSE
		\FOR{each $B_j \text{ assigned to } c_i$} \label{lin:3n}
		\IF{$buff(B_j, \theta)$ \textit{overlaps or encloses  A}} 
		\STATE $\mathcal{B}^n \leftarrow \mathcal{B}^n \cup \{nearby(A,B_j)\}$  \label{lin:4n}
		\ENDIF
		\ENDFOR
		\ENDIF 
		\ENDFOR
		\STATE \textbf{return} $\mathcal{B}^n$
	\end{algorithmic}
\end{algorithm}

Algorithm~\ref{alg:nearby} presents the pseudo-code for LD of relation `nearby'. Notice that the grid is constructed exactly as before, using the original areas in $\mathcal{B}$. In the pre-processing stage, the algorithm computes the mask $mask^{\theta}(c_i)$ for each cell using the $\theta$-buffered areas $B_j$ assigned to $c_i$. First, the cells $\Psi$ that enclose or overlap with  $A$ (i.e. the cells to which $A$ is assigned) are located (line~\ref{lin:1n}). For each cell $c_i \in \Psi$, if $mask^{\theta}(c_i)$ encloses $Overlap\_S(c_i,A)$ (line~\ref{lin:2n}), then  only disjoint relations to $B_j$ areas assigned to $c_i$ can be inferred. This is because of the way $mask^{\theta}(c_i)$ has been constructed. If this is not true, all areas $B_j$ assigned to $c_i$ need to be examined, and if $buff(B_j, \theta) $ overlaps or encloses $Overlap\_S(c_i,A)$, then we append with the discovered link $nearby(A,B_j)$ the result set $\mathcal{B}^n$ (lines~\ref{lin:3n}--\ref{lin:4n}). 


\subsection{Generalizing MaskLink to other Spatial Representations}

Apparently, as already described the above Algorithms~\ref{alg:topol} and~\ref{alg:nearby} can be  used for any type of spatial representation of $A$, namely point, poly-line or polygon. Furthemore, proximity relations between  points can be supported. In case any $B_i$ is a point, then $buff(B_i, \theta)$ is the area around that point, that is treated similarly to any other $\theta$-buffered area. However, this treatment of point-to-point relations concerns our future work. 
Another important feature of the proposed technique, is that it has minimum memory requirements, i.e., only the target data set needs to be organized in the grid, while the source data set is not necessarily a-priori known and accessible as a whole. Compared to state-of-the-art RADON, this is an important benefit since our proposed technique can be considered as memory-less with respect to the source data set, and it can be also applied in streaming data sources. Finally, the two algorithms can be assigned to different processing nodes, for further parallelization.



\section{Evaluation}\label{evaluation}

In this section, we report the results of our empirical study. MaskLink has been developed in Java 1.8. All experiments were conducted on a 6-core Intel(R) Xeon(R) CPU E5-2603 v4 @ 1.70GHz VM with 16GB RAM, running 64-bit JDK 1.8.0 on Ubuntu 16.04.2 LTS. Each experiment was assigned 8GB RAM and a timeout limit of 55 hours. Experiments which ran longer than this upper limit were terminated and reported as such in the experimental results.

\subsection{Experimental Setup}

The goal of the experimental study is twofold:
\begin{itemize}
	\item To demonstrate the efficiency of MaskLink compared to the state-of-the-art for topological link discovery (RADON~\cite{radon17aaai}), while being more generally applicable (Section~\ref{sub:expTop}).
	\item To quantify the performance gain achieved by the extension of MaskLink for proximity-based link discovery compared to the baseline approach, which is the only known solution to-date (Section~\ref{sub:expPro}).
\end{itemize}

For the first experiment on topological link discovery, we employ the data sets used in the evaluation of RADON~\cite{radon17aaai}, namely NUTS and CLC:
\begin{itemize}
	\item $\mathcal{A}=$CORINE Land Cover\footnote{see also https://datahub.io/dataset/corine-land-cover} (CLC) is provided by the European Environment Agency, which collects data regarding the land cover of European countries.	
	\item $\mathcal{B}=$NUTS\footnote{Version 0.91 (http://nuts.geovocab.org/data/0.91/)}, provided by Eurostat group of the European Commission,  contains a detailed hierarchical description of statistical areas for the whole Europe.
\end{itemize} 
\noindent Since CLC contains 44 data sets varying in size (from a few hundreds to hundreds of thousands), we merged all datasets into one big data set. For testing scalability, we exported from CLC data sets of varying size \{500K, 1M, 1.5M, 2M, 2.5M\} and evaluated MaskLink against RADON\footnote{downloaded from \url{https://github.com/dice-group/LIMES}, accessed on December 2017} and baseline. We also preprocessed the NUTS datasets, to convert the \texttt{ngeo:posList} serialization to Well-Known-Text format. All the reported experimental results do not include the preprocessing step. RADON  has been configured using the default settings, as in \cite{radon17aaai}. The MaskLink  and the baseline technique use a 2.5 degree  granularity grid. 

For the second experiment on proximity link discovery, we evaluate the proximity relation ``nearby'' using real-world datasets compiled from positions of vessels and fishing areas. Specifically, we use as $\mathcal{A}$ a data set that contains kinematic messages of vessels in the Mediterranean Sea spanning between 2016-01-11 to 2016-01-31, whereas $\mathcal{B}$ is a data set of fishing areas  that contains 5,076 polygons  generated from raster images depicting the fishing intensity in European waters (reported by European Union). The goal is to identify links between vessel positions and fishing areas that  these vessels approach.
We report results for different numbers of position messages \{500K, 1M, 1.5M, 2M, 2.5M\}, while using the complete data set of  fishing areas in the  Mediterranean Sea. 
Since RADON does not support point-to-region relations (the heuristic for deciding the size of cells based to the size of areas in input, cannot be applied), the MaskLink technique has been evaluated only against the baseline approach.

\subsection{Results for Region-to-Region Topological Relations} \label{sub:expTop}

We compare the MaskLink technique for all topological relations to RADON using the data sets CLC and NUTS.
RADON requires that the entire data sets are loaded in memory, which is not possible in our experimental setting. We overcome this memory limitation, by loading NUTS (as the smaller dataset) in memory, and accessing CLC in batches of lines. For a fair comparison of techniques, we repeat the same procedure for MaskLink,  although our technique can be directly applied on the given data sources. 

Table \ref{tbl:1} and the corresponding chart on its left report the total processing time of MaskLink for different number of CLC entries, in comparison with RADON and baseline. 
Specifically, the first column of the table indicates the size of CLC data set, and the next columns present the total processing time for RADON, MaskLink and baseline for computing all topological relations. 

We observe that the MaskLink consistently outperforms RADON by approximately 10\%. This is an important finding, as the main interest is the scalability of LD methods with input size. Also, recall that MaskLink achieves this performance using less memory and even when one data source is streaming. When comparing MaskLink to baseline, we observe that the gain achieved by MaskLink increases with the size of the data source, indicating that large instances of the problem cannot be efficiently solved by the baseline algorithm.

As far as the overhead imposed in construction due to the computation of the geometry of a mask is concerned, we observed that the time needed to compute the mask areas of all cells for the given target dataset was approximately 3 seconds. This is deemed tolerable, since it is a one-time cost and provides a considerable gain as shown by comparing the processing time of MaskLink to that of baseline.

\begin{table}[ht]
	\begin{minipage}[b]{0.50\linewidth}
		\centering
		\includegraphics[width=\textwidth]{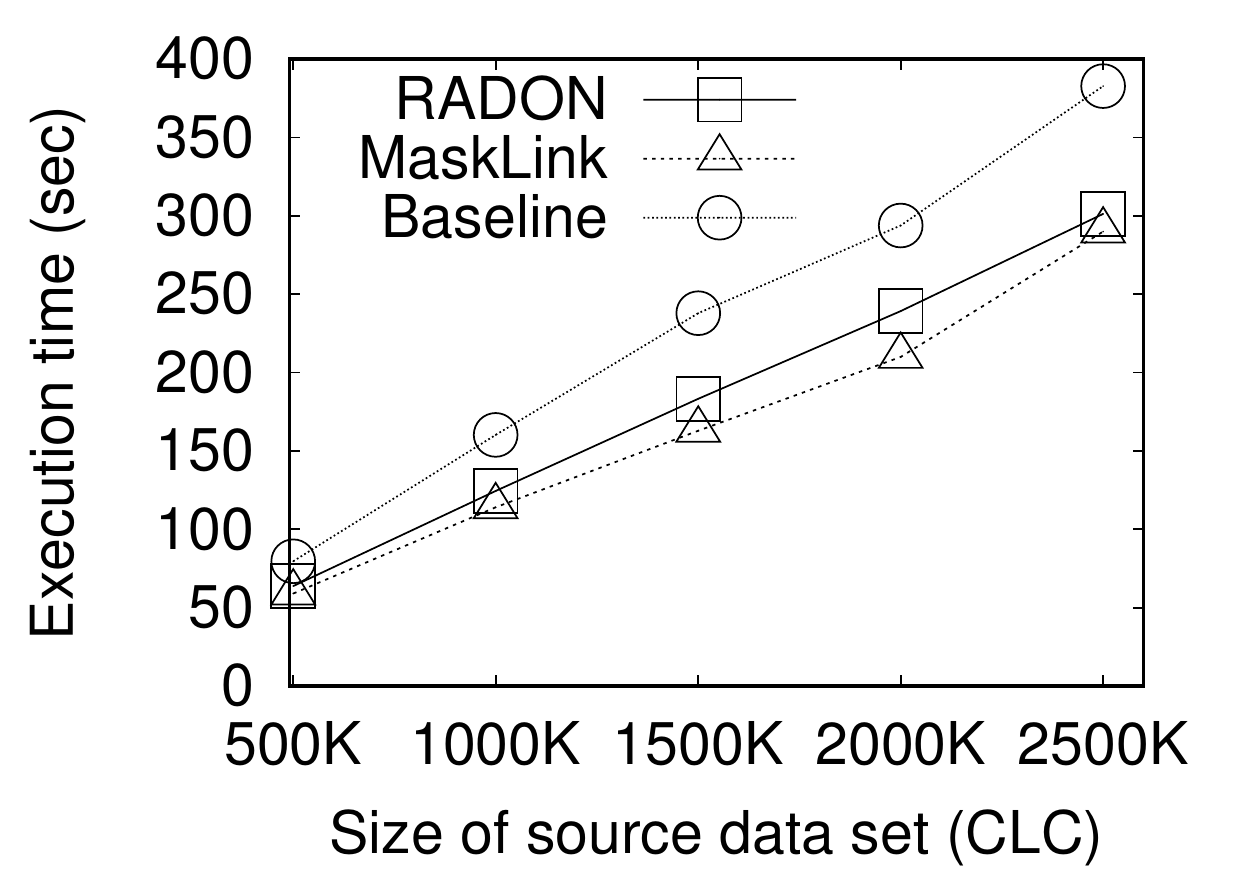}
		
		\vspace{32pt}
	\end{minipage}
	\vspace{12pt}	\begin{minipage}[b]{0.45\linewidth}
		\begin{tabular}{|p{1.7cm}|r|r|r|}
			\hline
			\#Entries in source (CLC) & RADON & MaskLink & Baseline \\ \hline
			500K & 63.812 & 59.05 & 79.65\\ \hline
			1000K & 124.549 & 114.05 & 160.28\\ \hline
			1500K & 183.359 & 162.99 & 237.89\\ \hline
			2000K & 239.312 & 210.00 & 293.84\\ \hline
			2500K & 301.289 & 289.97 & 382.84\\ \hline
		\end{tabular}
		\caption{Comparison of total processing time of RADON, MaskLink and baseline for all topological relations}
		\label{tbl:1}
	\end{minipage}
\end{table}

\subsection{Results for Point-to-Region Proximity-based Relation} \label{sub:expPro}


In this experiment, we report results for different number of position messages and using the complete data set of fishing areas in the Mediterranean Sea. Table \ref{tbl:2} and the chart on its left report the total processing time for all topological and proximity relations that can be computed between points and areas. 
The first column indicates the size of source dataset, while the second and third columns report the total processing time for MaskLink and baseline, respectively. 

Again, MaskLink outperforms the baseline consistently, but this time by a much larger margin, since the problem of proximity link discovery is harder in general. It is very significant to notice that the baseline is not scalable for this problem, as it does not terminate in reasonable time when the input size is larger than 1500K. In contrast, MaskLink scales gracefully with the size of input data. 

In the case of very small input sizes, the baseline is faster than MaskLink. This can be explained by the fact that for small datasets the cost of computing the mask of each cell becomes comparable to the total processing time. In any case, our focus is to improve the performance of LD for large data sets.

\begin{table}[ht]
	\begin{minipage}[b]{0.50\linewidth}
		\centering
		\includegraphics[width=\textwidth]{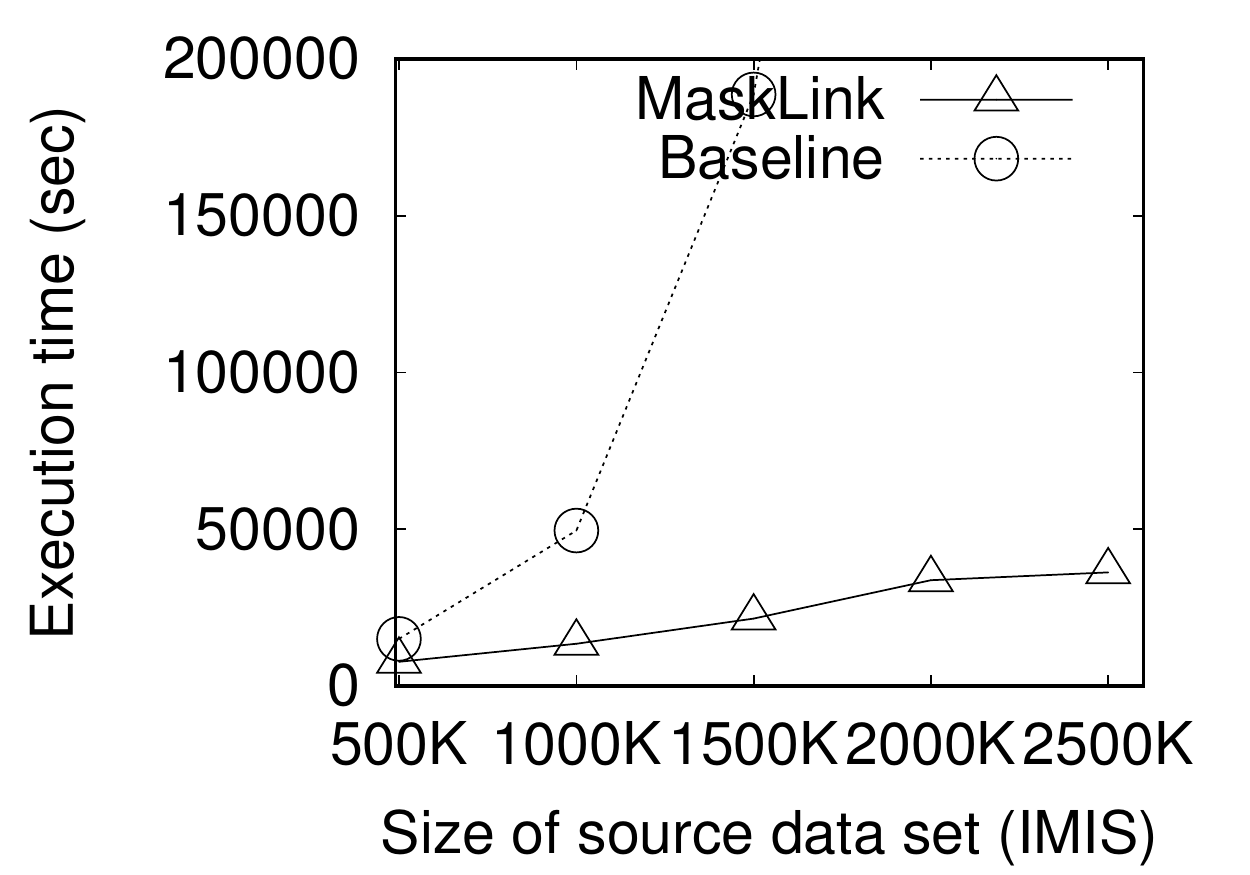}
		
		\vspace{28pt}
	\end{minipage}
	\begin{minipage}[b]{0.45\linewidth}
		\begin{center}
			\begin{tabular}{|p{2cm}|r|r|r|}
				\hline
				\#Entries in source (IMIS) & MaskLink & Baseline\\ \hline 
				500K & 7769.05 & 15064.83\\ \hline 
				1000K & 13494.58 & 49630.11\\ \hline 
				1500K & 21528.29 & 188763.63\\ \hline 
				2000K & 33801.71 & timeout\\ \hline 
				2500K & 36275.21 & timeout\\ \hline 
			\end{tabular}
			\caption{Total processing time for topological and proximity relations between points and areas using MaskLink and baseline.}
			\label{tbl:2}
		\end{center}
	\end{minipage}
\end{table}

Furthermore, as a side-note, we observed in this experiment that the time required for preparing the target dataset (i.e., populating the grid and computing the mask) is considerably larger than the time needed for the first experiment. This is explained by the fact that the geometries used in the second experiment are produced from raster images (vectorized), thus the geometries have more fine-grained detail and the mask computation requires more processing time.

\section{Conclusions} \label{sec:concl}
This paper presents a filtering technique for link discovery of topological and proximity relations considering also streaming data sources. To the best of our knowledge, currently, there is no other work that can efficiently compute both topological and proximity relations. 
The technique is based on the observation that it is common fact that some space in the cells of space-tiling blocking is never covered by spatial representations. Therefore, for a given spatial entity, we can detect the corresponding cell for that entity, and evaluate against the geometry defined by the unused space in that cell. Thus we avoid the comparisons to all the rest spatial entity already assigned to the cell, if the given entity is enclosed in the unused area. This technique can be also applied to higher dimension space-tiling (e.g. 3D), it is memoryless since it does not employ optimizations or heuristics that need to be computed from accessing the entire datasets. Further more, since no a-priori knowledge of the datasets is needed, this technique can be efficiently used on streams of data, which is in planned for next tasks.

\vspace{12pt}
\small
\noindent\textbf{Acknowledgements:} This work is supported by the datAcron project, which has received funding from the European Union's Horizon 2020 research and innovation program under grant agreement No 687591.
\normalsize

\bibliographystyle{abbrv}           
\bibliography{bibfileLD}        

\end{document}